# Substrate Dependent Resistive Switching in Amorphous-HfOx Memristors: An Experimental and Computational Investigation

*Pradip Basnet[†], Darshan G Pahinkar[‡], Matthew P. West[†], Christopher J. Perini[†], Samuel Graham[‡] and Eric M. Vogel[†]*

[†]School of Materials Science and Engineering, Georgia Institute of Technology, Atlanta, Georgia 30332, USA
[‡]Electronics Manufacturing and Reliability Laboratory, Department of Mechanical Engineering, Georgia Institute of Technology, Atlanta, Georgia 30332, USA
Address correspondence to: pradip.basnet@mse.gatech.edu

**ABSTRACT**

*While two-terminal $HfO_X$ (x<2) memristor devices have been studied for ion transport and current evolution, there have been limited reports on the effect of the long-range thermal environment on their performance. In this work, amorphous-$HfO_X$ based memristor devices on two different substrates, thin $SiO_2$(280 nm)/Si and glass, with different thermal conductivities in the range from 1.2 to 138 W/m-K were fabricated. Devices on glass substrates exhibit lower reset voltage, wider memory window and, in turn, a higher performance window. In addition, the devices on glass show better endurance than the devices on the $SiO_2$/Si substrate. These devices also show non-volatile multi-level resistances at relatively low operating voltages which is critical for neuromorphic computing applications. A Multiphysics COMSOL computational model is presented that describes the transport of heat, ions and electrons in these structures. The combined experimental and COMSOL simulation results indicate that the long-range thermal environment can have a significant impact on the operation of HfOx-based memristors and that substrates with low thermal conductivity can enhance switching performance.*

**INTRODUCTION**

Metal Oxide (MO) based resistive random access memory (ReRAM) devices have been drawing a lot of attention recently for their various potential applications, including neuromorphic computing.[1-3] Many researchers have demonstrated that a hybrid system, composed of complementary metal oxide semiconductor (CMOS) neurons and memristor synapses, can support important neuronal functions such as spike timing dependent plasticity (STDP).[4, 5] Wang et al. have reported that a diffusive memristor and its dynamics can accomplish a direct emulation of both short- and long-term plasticity of biological synapses, while other researchers have established MO based synapses at the device level.[6, 7]

Among the typical binary transition MOs (e.g. $TiO_2$, $TaO_2$, $ZrO_2$ and $HfO_2$), $HfO_X$ (x<2; sub-stoichiometric) is very popular as a valance change memory (VCM) material due to its promising physical and chemical properties.[2, 8, 9] Oxide-based VCMs are one type of memristive device, where changes in the resistance or conductance can be induced by redox reactions in the oxide within a filament or at the oxide–electrode interfaces under DC bias voltage or a given voltage pulse.[10, 11] $HfO_X$ is both chemically and physically stable and also compatible with CMOS technology.[9] Amorphous (a)-$HfO_X$ offers a few additional advantages, such as high uniformity over large area, low temperature synthesis and controlled electrical properties.[12] To date, both amorphous and poly-crystalline stoichiometric and sub-stoichiometric (oxygen deficient) $HfO_X$ based memristors have been studied experimentally and computationally,[12-18] and more importantly, stand-alone synaptic devices have been demonstrated.[19] While there has been significant prior research related to digital switching in filamentary adaptive $HfO_2$, there has been a very limited fundamental understanding related to the primary factors controlling analog resistance change.[3, 20] The analog synaptic properties of $HfO_X$ such as multilevel set and reset states need further investigation. It is worth noting that biological synapses show analog behavior with multilevel synaptic weight changes, and analog conductance (or resistance) change of two-terminal MOs based memristor devices is quite similar.[3] Recent research efforts along these lines have made a lot of progress. For example, Long et al.[8] have observed the sharp and gradual reset process in unipolar switching in $HfO_2$, while a compliance-free, digital set, and gradual-reset has been reported in $TaO_x$ by Abbas et al.[10]

Forming a conductive filament (CF) in the MOs involves a creation of oxygen-deficient (or metal rich) region that requires breaking metal-oxygen bonds followed by migration of oxygen anions and/or vacancies; both of these processes are controlled by the local temperature field and the applied electric field.[1, 3, 21] As a result, a "filamentary based" memristor operates at extremely harsh conditions of electric field up to 10 MV/cm and current density ~1 MA/cm$^2$ or above.[13] It is well accepted that bipolar device behavior is controlled by the motion of negatively charged oxygen ions (or the positively charged oxygen vacancies) through a conducting filament (CF).[22, 23] However, very little is known about how the thermal boundary materials (TBMs) affect the oxygen ions' drift, diffusion, and CF evolution process.[24-26] The main difficulties in understanding the device behavior is that the filament has been measured to be on order of a nanometer in diameter and re-oxidation of the filament tip only occurs about a nanometer in distance from the oxygen reservoir near the top electrode (TE) as shown in Fig.1 (top).[16, 17, 26-28] Local temperature rises as high as 1000 K have been reported, which depends on the thermal conductivities of the oxide layer, filament, electrodes, and substrate.[14, 25] Since both the drift and diffusion of the oxygen ions and/or vacancies can occur simultaneously, it is very difficult to quantify the digital or analog switching behavior that can occur on the nano-second timescale.[25, 29] A variety of approaches, including computational and experimental methods, have been used to explain fundamental processes association with vacancy/ion migration at the metal-oxide interfaces.[14, 17, 30, 31] Syu et al. showed that the high-speed resistive switching (RS) behaviors of a memristive device is due to only a few atoms involved in the redox reactions and mixed ionic-electronic transport at the interface, which makes experimental study even more complicated.[29] Recently, Kim et al. studied TaO$_x$ device behavior using three different TEs, namely: Pd, Ru, and W, with thermal conductivity values in the range of 71.8 to 173 W/m-K, and reported the effect of the TBMs on overall resistive switching.[26] However, the use of different TEs to study the effect of TBMs can be misleading as other properties of the top electrode, like its work function, oxygen affinity etc., can heavily affect the RS. Although TiO$_2$ and HfO$_2$ based memristors have been fabricated on flexible (e.g. plastic) substrates,[12, 32] a systematic study of the impact of the thermal conductivity of the substrate has not been performed.

In this work, the performance of 5 nm sub-stoichiometric a-HfO$_x$ (x ≈ 1.8) based memristor devices, fabricated on either thin SiO$_2$(280nm)/Si or glass substrates, are compared and reported. The thin SiO$_2$(280nm)/Si substrate has an effective thermal conductivity of 1.2 W/m-K, whereas the glass substrate has a thermal conductivity of 138 W/m-K. Experimental results for electroforming, digital (abrupt) set, analog (gradual) reset and endurance were compared. A Multiphysics COMSOL model that simulates the simultaneous drift, diffusion and thermophoresis of oxygen vacancies was used to estimate the local and temporal temperature profile and its effect on the overall device performance. The comparison of current-voltage (I-V) curves from the COMSOL model and those from the experiments provide the validity of the chosen approach in these models. Our combined experimental and modeling findings provide critical insights into the impact of the long-range thermal environment on memristor operation.

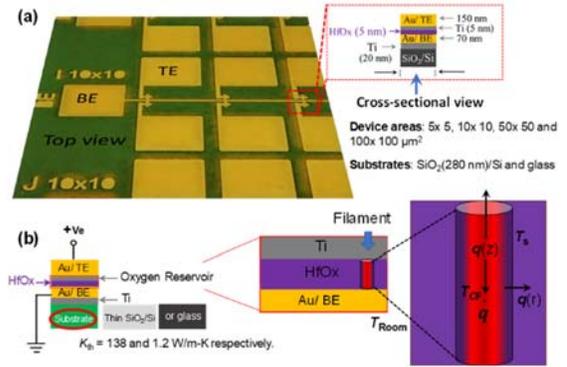

Fig. 1 (a) Schematic of the as-prepared Au (BE)/HfO$_x$/Ti/Au (TE) devices, 2D views: (left) Top- and (right) cross-sectional views. (b) A typical electrical characterization set up and the process of filament formation: (left to right) substrates thermal properties, device layers, and nanoscale filament formation in HfO$_x$. Arrows showing the heat transfer directions, namely: axial q(z) and radial q(r), from the filament to the surroundings. Note: drawings are not to scale.

**RESULTS AND DISCUSSION**

Prior to characterizing the resistive switching behavior, it is important to first demonstrate that the substrates did not impact the intrinsic structure and behavior of the materials and devices. Figs. 2 (a) and (b) show the capacitance and initial resistance of the devices, of different areas in the range: 5×5 to 100×100 μm$^2$, measured prior to forming the devices. Results show that the measured capacitance (at 10 KHz) increases linearly as a function of device area but is independent of substrate for a given area. For example, the capacitance of both the devices on thin SiO$_2$/Si and glass substrates of sizes 100 and 10,000 μm$^2$ are measured to be ~ 5.0 and ~500.0 pF, respectively.

These results agree well with the thickness of oxide layer (~ 5 nm) while considering the devices as a parallel plate capacitor, whose capacitance can be estimated from: $C = (\varepsilon_o \varepsilon_{ox} A)/t_{ox}$; where $\varepsilon_o$, $\varepsilon_{ox}$ and $t_{ox}$ are the permittivity of vacuum (= 8.85×10$^{-12}$ F.m$^{-1}$), dielectric constant (HfO$_x$ ~ 20) and thickness of the oxide layer, respectively. Similarly, the initial device resistance is substrate independent for the given area, A = 10 × 10 μm² (Fig. 2(b)). The capacitance and initial resistance results, therefore, confirm that the intrinsic electrical device behaviors are substrate independent. These results strongly suggest that any differences in forming and switching behavior are not related to differences in the structure of the active materials.

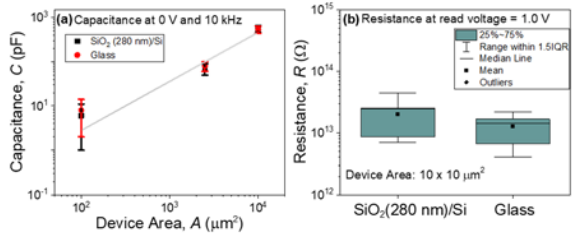

Fig. 2 Electrical properties of the Au (BE)/a-HfO$_x$/Ti/Au (TE) pristine devices, i.e. before electroforming, deposited on SiO$_2$(280 nm)/Si and glass substrates: (a) Capacitance of three different size devices and (b) Resistance of 10 ×10 μm² size devices as indicated. The error bars are obtained from 18 devices of each specified.

Fig. 3(a) shows representative results for forming, reset, and set cycles of a device on the thin SiO$_2$/Si substrate. The resistance decreases by approximately nine orders of magnitude from point A to point E (~$10^{13}$ Ω to ~$10^4$ Ω). The current from point A to point E at ~3.5 V is small (<$10^{-11}$ and associated with the pristine device behavior without a filament. From ~3.5 V to ~4.0 V (point B to point C), there is a formation of an initial filament resulting in a significant increase in the current by approximately two orders of magnitude. From point C point D (~4.0 to 4.6 V in this example), the current fluctuates. This may indicate that there is a competition of ion-electron migration due to thermal and electrical fields resulting in an increase or decrease in the filament number or size. This step plays a critical role in determining the overall resistance or size/number of the filament(s). Statistical result shows that the forming voltage ($V_f$) for both thin SiO$_2$/Si and glass substrates is approximately the same as shown in Fig. 3(b). However, the filament resistance after ~3 V is substrate dependent (see Electronic Supplementary Information, †ESI) and it could be attributed to the local heating near the beginning of filament formation (say, near point D). To make a fair comparison of the RS, a stabilization process is performed after the filament formation, as described in detail in the †ESI (S1 to S3).

Fig. 4(a) shows the on state device resistance ($R_{on}$) immediately following filament formation and stabilization. The glass devices show lower $R_{on}$ than the Si/SiO$_2$ devices suggesting that the low thermal conductivity substrates assist forming either wider or more filaments. The overall process of growing wider/more filament(s) could be due to a thermally assisted mechanism during the filament formation or stabilization process. Figures 4(b), (c) and (d) show the set and reset hysteretic I-V characteristics for maximum reset voltages of -1.0 V, -1.5 V and -1.8 V, respectively. The current following set (reset) for the glass devices is higher (lower) than the Si/SiO$_2$ devices. Furthermore, the tripping voltage (the voltage at which the device begins to reset) is smaller for the devices on glass as compared to SiO$_2$/Si. This earlier tripping voltage and smaller LRS (lower resistance state) for the case of glass substrate may be attributed to presence of larger filaments as compared to SiO$_2$/Si substrates, resulting in a higher current and filament temperature. This means that the filament re-oxidation begins at a lower voltage resulting in a larger HRS (high resistance state), and larger memory window (MW) for the glass devices. A wider HRS to LRS ratio (high to low resistance ratio, HLR) of the glass devices, i.e. in the range: 4 to 100 with |$V_r$| = 1 to 1.8 V (see Fig. S4(b) in the †ESI) confirmed the better performance of the devices on glass.

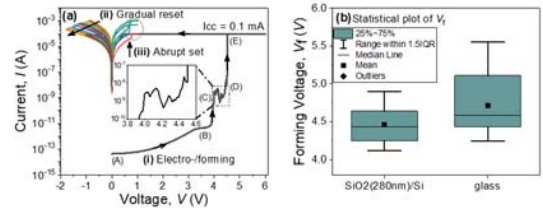

Fig. 3 (a) Electroforming (/or forming; curve (i): (A) to (E)), initial reset (ii) and set (iii) steps of a pristine Au (BE)/a-HfO$_x$/Ti/Au (TE) device, fabricated on a thin SiO$_2$(280)/Si substrates. Note that the three steps i) to (iii) shown in Fig.(a) helps getting stable conductive filament/s and stable resistive switching. (b) Comparison of substrate dependent forming voltage, $V_f$ of 24 devices from each SiO$_2$(280)/Si and glass substrate.

The digital reset/set of the devices is well explained in the literature with the help of atomically thin barrier layer that is created/destroyed during the reset/set processes.[29] However, there is still lack of a detailed mechanism for the smooth analog reset response of the HfOx based memristor. A more detailed mechanism will be discussed later, in the next section, with the help of a Multiphysics COMSOL simulation results.

Fig. 5 shows the representative multi-level resistance at various set and reset voltages, each of 100 test cycles for the devices on both thin SiO$_2$/Si and glass substrates. Results show that all the devices exhibit multi-level resistance (or conductance) values at some appropriate applied voltages. Also, it is evident that the devices on glass substrate retain more distinguishable HRS and LRS values for the same range of set and reset voltages, and they can function without a visible degradation in the operating voltages as shown in Fig.5(a). In other words, the devices on glass substrate exhibited clearly distinguishable MWs at lower reset voltages (starting from -0.8 V, in our case), whereas no such reliable separation of LRS and HRS was observed with the devices on SiO₂/Si substrate (results not shown here). As expected, the LRS values of the glass devices are smaller than that of the devices on SiO₂/Si substrate, while the HRS values are greater (*see* Fig.5(a)-(b)).

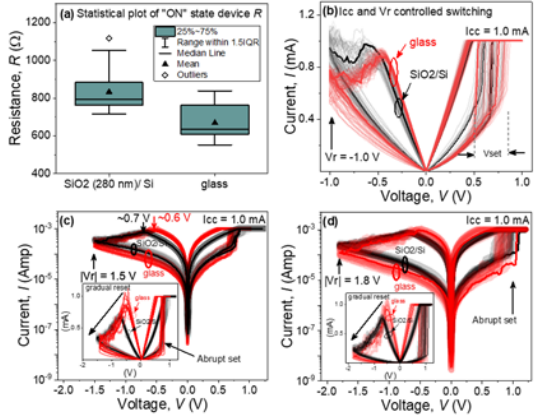

Fig. 4 Substrate dependent resistive switching performance comparisons of the Au/a-HfOₓ/Ti/Au (TE) devices: (a) Initial set or ON state device resistance after forming and stabilizing the filament/s and (b)- (d) digital (abrupt) set and analog (gradual) reset switching starting with the same compliance current, Icc = 1.0 mA, but different reset voltages, |Vr| = 1.0 to 1.8 V.

The difference in HRS values becomes more prominent at higher $V_r$, resulting the higher HLR values of the glass devices as observed in the digital response. The observed low operating voltages and the multiple HLR values especially of the devices on glass substrate is consistent with the literature of sub-/stoichiometric HfOₓ based memristors.[12,13,24,33,34] The devices on glass substrates exhibited noticeably improved MW, which agree well with the results from digital switching. The enhanced performance of the devices on glass substrate can be attributed to the fact that both the formation of CF/s and reset switching were favorable with the low heat dissipation from the substrate to the environment. However, it is important to understand that the different filament sizes (and perhaps geometry) for different substrates must be verified with high resolution in-situ transmission electron microscopy (TEM) studies.

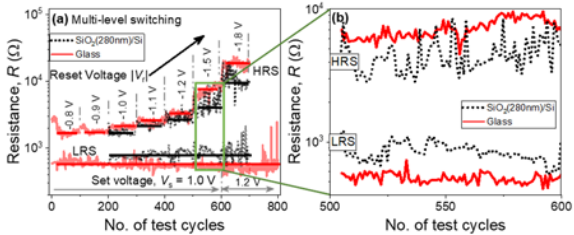

Fig. 5 Substrate dependent analog resistive switching of Au/a-HfOₓ/Ti/Au (TE) devices: (a) Effect of incremental applied reset voltages (|Vr| = 0.8 to 1.8V), showing multi-level resistive switching starting with different set and reset voltages. (b) (Zoom-in) Showing the performance of glass and SiO₂/Si devices at 1.0 V (set) and -1.5 V (reset) voltages.

**Thermal modeling and experimental validation.** The substrate dependent RS performance of the devices with stable CF of known initial size and geometry are discussed in greater details in this section, followed by the experimental results validation. The basic hypothesis of this thermal modeling is that the RS switching behavior of a filamentary device is controlled by both the applied electric field (***E***) and local thermal field, generated via joule-heating. For this, a COMSOL physical model that is reported in our recently published article[35] has been used to analyze the intermediate variables such as local temperature and oxygen defect (vacancy) concentration, so that their role in the device performance can be understood. The thermal model considers the drift, diffusion and thermophoresis of oxygen vacancies that are instrumental in creating and suppressing the CF. The vacancy conservation that includes these phenomena is solved in conjunction with current conservation and energy conservation equations as described in Eq. (1) through (3).

$$\frac{\partial n_V}{\partial t} + \nabla \cdot (v_V n_V) = \nabla \cdot (D_V \nabla n_V) + \nabla \cdot (S_V D_V n_V \nabla T) \quad (1)$$

$$\nabla \cdot \left( \sigma \nabla \psi + \varepsilon \frac{\partial}{\partial t} \nabla \psi \right) = 0 \quad (2)$$

$$\frac{\partial T}{\partial t} = \frac{1}{\rho c} \nabla \cdot (k \nabla T) + \frac{\sigma}{\rho c} (\nabla \psi)^2 \quad (3)$$

In these equations, $n_v$ is the density of oxygen vacancies, $v_v$ is the drift velocity of vacancies, $D_V$ is the diffusion coefficient of the same, $S_V$ is the thermophoresis coefficient, $\sigma$ is the electrical conductivity, $\psi$ is the voltage potential, $\varepsilon$ is the permittivity of the oxide material, $T$ is the temperature, $\rho$ is the density, $c_p$ is the specific heat, and $k$ is thermal conductivity. Finally, the validity of experimental results is confirmed by taking two different radii of CFs, namely: 3 nm for SiO₂/Si and 4 nm for glass substrates, which are more realistic than taking a single size filament as discussed above. Further details of the model are not described in this article for brevity and can be found in the work by Pahinkar et al.[35]

Fig. 6 shows a comparison of I-V curves from the experiments and the model simulations demonstrating a good agreement, especially on the reset side. The earlier tripping voltage of the glass device further confirm the validity of this model: Bigger or wider CF of glass device, with lower initial resistance, utilizes more power (= $V^2/R$) forcing the CF to rapture earlier or vice versa. The simulation results closely follow the substrate dependent trend for both reset voltage and the MW. However, further refinement of the model is necessary for better curve fitting over the whole I-V range. The faster decline in the current of the glass device after the tripping voltage clearly represents higher HRS compared to the SiO$_2$/Si device. The difference in the two slopes of the I-V curves, namely: from -0.5 V to -1 V and -1 V to 0 V, may indicate that the device reset switching is not only controlled by the CF size but also the substrate's thermal properties. Thus, the observed higher HRS and MW of glass devices can be attributed to the negligible heat loss, resulting an earlier rapture and re-oxidation of the CF/s. Fig. 7 sheds light on how the insulating or dielectric barrier is created during the reset stage and breakdown of the barrier takes place during the digital set stage. Because negatively charged oxygen ions move from the capping layer Ti into the filament region and a barrier layer, close to the BE, is created at the end of the reset stage as shown in Fig. 7(a). This barrier layer has a lower electrical and thermal conductivity than the region elsewhere and becomes the bottleneck for the current flow. This layer also increases the local temperature gradient, it being the region of the high potential drop. Thus, an HRS is achieved. When the polarity is reversed to execute the set stage, the temperature increases initially with the current in the off-state and when the temperature reaches a threshold after which the vacancies are mobile, drift and diffusion acting the in the same direction nearly instantaneously cause the breakdown of the barrier layer as seen in Fig. 7(b). It can also be seen that the oxygen ions enter the Ti layer after set, which can be inferred from the fact that the number of vacancies in the Ti layer have dropped. This is consistent with the observed abrupt set for all devices. In conjunction to this, a continuous metal rich filament reestablishes in the HfOX layer at the end of the set stage.

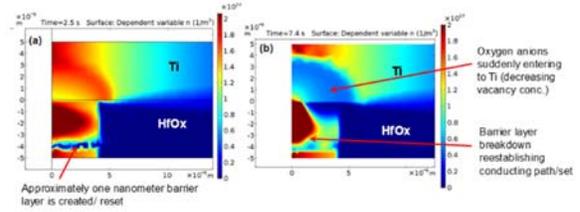

*Fig. 7 (Color online) Oxygen vacancy distribution (a) at the end of the reset stage (b) At the end of the set stage for glass substrate.*

of TE temperature, as a function of voltage, respectively. When the simulations for glass and thin SiO$_2$ (280 nm)/Si substrates begin with the same initial conditions, the temperature of both the filament tip and the TE in the glass substrate is marginally higher than that in the SiO$_2$/Si substrate. This is clearly attributed to more heat remaining near the device for the same power (the current and voltage are the same for both substrates), because of the glass thermophysical properties. This means that the temperature required for onset of the migration of vacancies is achieved earlier in the glass device than that in the SiO$_2$/Si device. As the migration begins creating the dielectric barrier, the current begins to drop lowering the temperature as we observed from the experiments (*see* Fig. 6). Although, the difference in the temperature profiles of filament tip in the set side are not significant for our devices, the impact of substrate material on the TE temperature is more tangible. As seen from Fig. 8(b), the temperature rise in the glass substrate to be more than 40 °C, while that in the SiO$_2$/Si substrate is less than 5°C. This behavior is obviously due to different fractions of thermal power going upward toward the TE and downward toward the substrate for different substrates. The glass substrate provides more resistance to heat transfer than SiO$_2$/Si, hence more fraction of the total thermal power in the filament goes toward the TE for glass, thereby showing an appreciable temperature rise.

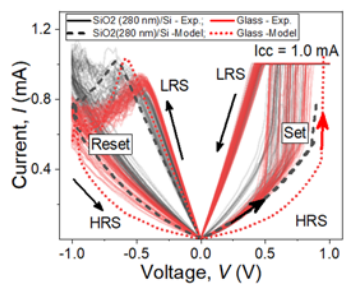

*Fig. 6 Substrate dependent comparison of I-V plots for thin SiO$_2$(280 nm)/Si and glass substrates obtained from the experiments and COMSOL model simulations. Note that different sizes of conducting filaments (CFs), namely: 3 and 4 nm in diameter, were used in the modeling for SiO2/Si and glass substrates respectively.*

The role of substrates dependent temperature field on the defect concentration in HfO$_X$ or the oxygen vacancy activation and movement resulting different RS performance was also studied. Fig. 8(a) and (b) show the comparison of the filament tip temperature (at the interface between Ti and HfO$_X$) and the exposed surface

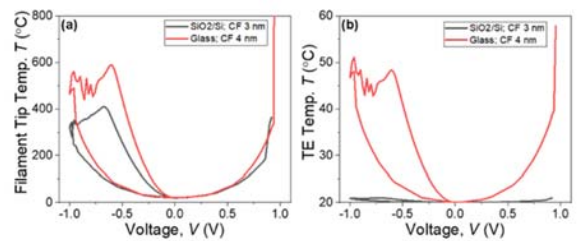

*Fig. 8 Substrate dependent temperature of (a) filament tip (b) top electrode (TE) surface, plotted with the applied voltage for thin SiO$_2$/Si and glass substrates. Note that 3 and 4 nm of conducting filaments (CFs) were used in the COMSOL modeling for SiO2/Si and glass substrates respectively.*

Finally, the models were simulated to confirm that the factors considered in the present model that affect the oxygen anions and/or vacancies migration can indeed reproduce the non-volatile nature of the 'memory' or internal resistance of the device. Fig. 9 shows the typical gradual reset I-V curves for a series of |$V_r$| in the range: 0.5 V to 1.0 V for a glass substrate. The ion migration and barrier layer creation at every voltage step is remembered by the device perfectly validating the approach considered, when compared with the experimental results (*see* Fig. S3(a)). To the best of our knowledge, this type of modeling of non-volatile reset has not yet been reported in the literature. The use of this modeling approach to further validate nanoscale ion transport phenomena as a function of current flow and temperature field for $HfO_X$ and other MOs exhibiting similar properties is thus warranted and ongoing.

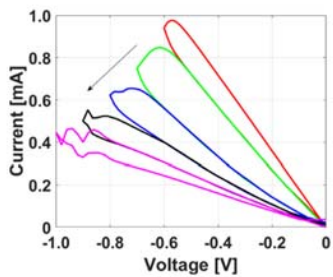

**Fig. 9** *The I-V curves from the COMSOL model showing gradual reset with a non-volatile resistance memory.*

**CONCLUSION**

In summary, Au (BE)/a-$HfO_X$(5 nm)/Ti/Au (TE) memristor devices were fabricated on two different substrates, thin $SiO_2$(280 nm)/Si and glass, and characterized to study substrate dependent device behavior. The reported memristor device fabrication technique is simple, scalable, and highly reliable (almost 100% device yield in our case). Our study showed that the bipolar RS, both digital and analog, of the devices is dependent on the thermal diffusivities of the substrates. Devices on glass substrates showed the best RS switching performance, including a wider MW and a lower reset voltage, than the devices on $SiO_2$/Si substrates. This is attributed to the low thermal conductivity of the glass, enabling the device to hold the temperature and hence improve performance due to favorable diffusion of oxygen ions at the $HfO_X$ -Ti interface. Comprehensive computational modeling results confirm that the oxygen ions move back and forth between the Ti "capping layer" and the $HfO_X$ "active oxide layer" during reset and set stages. The movement of ions is reported as a primary function of applied electric field, thermophoresis, and diffusion, while the kinetics are governed by the material properties of the substrates. Our results indicate that the a-$HfO_X$ based memristor devices can be a promising candidate for neuromorphic computing. However, for commercial production, besides the requirement of device size and behavior, other physical parameter and limitations such as compatibility of substrates, power consumption, device packaging *etc.*, could affect the RS switching, and therefore, it may require an extensive study to further improve the analog response of these devices.

**Experimental Section**

   **Device fabrication.** All devices were fabricated on about 1"x 1" square size cleaned substrates. The Si (Orient: <100>, test grade) wafer were procured from University Wafer, and the surface was thermally oxidized to create ~ 280 nm thick $SiO_2$, using dry oxidation at 1000 °C. While the microscope glass slides (Gold Seal Catalog No. 3010) were used as received. The sneak current path free devices of different sizes were fabricated for different experiments (details are presented in the result section). Prior to spin coating the substrates with negative photoresist, NPR (NR-71 3000P), we cleaned the substrates with organic solvents and piranha solution, then dried with $N_2$ blower. As described in our recent paper, both the bottom and top electrodes (BE and TE) were patterned using a maskless aligner (Heidelberg Instruments MLA 150),[36] followed by the metal depositions using e-beam evaporation (20 nm and 5 nm Ti underneath the 70 nm BE and 150 nm TE respectively, deposited at the fixed deposition rates 1.0 Å/s). Both the BE and TE layers were deposited using 99.999% Au source at ultra-high vacuum (~3x$10^{-6}$ Torr) and the Ti layers were deposited without breaking vacuum using 99.95% Ti source pellets (Kurt j. Lesker). Note that the 20 nm adhesion layer of Ti film was deposited to improve the adhesivity of Au layer, while the 5 nm Ti was used as a buffer layer (/or oxygen getter capping layer) to improve the oxygen diffusion, as described in the literature.[5, 31] The $HfO_X$ (~ 5 nm) was synthesized using atomic layer deposition (ALD), Cambridge Nanotech Fiji F202 system. Thermal ALD process was used at 250 °C, with the Tetrakis (dimethylamido) hafnium (TDMA-Hf) and DI water as oxygen precursors. Before depositing the $HfO_X$ film on the actual device's substrate, the deposition rate was confirmed with a test run to ensure the accuracy for the final thickness and used the optimized rate to estimate total number of cycles needed (*i.e.* 45 cycles at 1.12 Å/cycles for 5 nm). Then, the final thickness and the uniformity of the $HfO_X$ films were also confirmed with the multi-point thickness measurements using an ellipsometer (J.A. Woollam M2000), measured to be 5.0 ± 0.1 nm (with uniformity 2%). After the BE and TE deposition, lift-off processes were performed with acetone at room temperature for a period of at least 8 hours. Also, the samples were sonicated for about 30 secs before the final rinse off to remove the PR completely. Finally, the HfO$_X$ film, outside the crossbar region, was etched by reactive ion etch (RIE) starting with SF6/Ar mixture (flow rates: 30/5 sccm) for a minute, setting the RF power at 200 W and the pressure at 40.0 mTorr. It is noteworthy that the as-prepared devices were characterized for the electrical properties without any post-fabrication treatments.

**Materials and Device Characterization.** As-deposited HfO$_X$ films were characterized for some of the important physical and chemical properties that are required, or at least helpful, to investigate the device RS behavior. For example, amorphous phase was confirmed with glancing angle X-ray diffraction (XRD) measurements (PANalytical Materials Research Diffractometer, MRD), with Cu K$\alpha$ ($\lambda$ = 1.540598Å) and 2$\vartheta$ in the range from 20°-80°, at a step size of 0.01°. And, X-ray photoelectron spectroscopy (XPS) was performed for chemical composition using a Thermo Scientific K-alpha XPS system with a monochromated Al K$\alpha$ X-ray source and a hemispherical analyzer in constant analyzer energy mode (50 eV pass energy, 0.1 eV step size, and a beam size of 400 μm). From the XPS analysis, the Hf:O ratio was estimated to be 1:1.8, which corresponds to a stoichiometric amorphous HfO$_{1.8}$.

**Electrical characterization**. All electrical measurements were performed using a Keithley 4200-SCS (Semiconductor Characterization System) and a Cascade Microtech probe station at room temperature and ambient conditions. As reported in our previous work, we characterized the devices with and without an external transistor to confirm if the transistor is required to avoid the possible damages to the devices: For example, permanent damage of the conductive filament or the performance degradation.37 Bias was applied to the TE while the BE was grounded. For DC sweeping voltage ramping was also fixed (= ±0.02 V) for all measurements in both positive and negative potential windows. Initial device characteristics were compared by measuring the capacitance-voltage (C-V) and initial resistance of the pristine devices, of all four different sizes as mentioned above. Then, the switching performance of 10 x 10 μm2 size devices, fabricated on both thin SiO2(280 nm)/Si and glass substrates, were measured and analyzed. Please note that all the test devices, on both glass and SiO2/Si substrates, were characterized without using the thermal paste; and this can cause a slight change in the device behavior that might be not easy to quantify.

*Notes*: The authors declare no competing financial interest.

*Acknowledgement*. The authors acknowledge Professors William A. Doolittle and Wei-Cheng Lee for the helpful discussion. This work is supported by the Air Force Office of Scientific Research MURI entitled, "Cross-disciplinary Electronic-ionic Research Enabling Biologically Realistic Autonomous Learning (CEREBRAL)" under award number FA9550-18-1-0024. This work was performed in part at the Georgia Tech Institute for Electronics and Nanotechnology, a member of the National Nanotechnology Coordinated Infrastructure, which is supported by the National Science Foundation (grant ECCS1542174).

**†*Electronic Supplementary Material (ESI)***: Substrate dependent conductive filament (CF) formation; the CF/s stabilization; the current–voltage (I–V) curves of stabilized CF/s; substrate dependent switching with the stable CF/s at different reset voltages; and substrate dependent power consumption. This material is available free of charge *via* the Internet at http://www.rsc.org.